\documentclass[12pt]{article}
\usepackage[latin1]{inputenc}

\usepackage{amsmath}
\usepackage{amsfonts}
\usepackage{amssymb}
\usepackage{graphicx}
\usepackage{geometry}
\usepackage{amssymb,epsfig,subfigure}
\usepackage{hyperref}

\makeatletter
\renewcommand\section{\@startsection {section}{1}{\z@}%
                                 {-3.5ex \@plus -1ex \@minus -.2ex}
                                   {2.3ex \@plus.2ex}%
                                   {\normalfont\large\bfseries}}
\renewcommand\subsection{\@startsection{subsection}{2}{\z@}%
                                   {-3.25ex\@plus -1ex \@minus -.2ex}%
                                     {1.5ex \@plus .2ex}%
                                     {\normalfont\bfseries}}
\renewcommand\subsubsection{\@startsection{subsubsection}{3}{\z@}%
                                   {-3.25ex\@plus -1ex \@minus -.2ex}%
                                     {1.5ex \@plus .2ex}%
                                     {\normalfont\itshape}}
\makeatother

\def\pplogo{\vbox{\kern-\headheight\kern -29pt
\halign{##&##\hfil\cr&{\ppnumber}\cr\rule{0pt}{2.5ex}&\ppdate\cr}}}
\makeatletter
\def\ps@firstpage{\ps@empty \def\@oddhead{\hss\pplogo}%
  \let\@evenhead\@oddhead 
}
\def\maketitle{\par
 \begingroup
 \def\thefootnote{\fnsymbol{footnote}}
 \def\@makefnmark{\hbox{$^{\@thefnmark}$\hss}}
 \if@twocolumn
 \twocolumn[\@maketitle]
 \else \newpage
 \global\@topnum\z@ \@maketitle \fi\thispagestyle{firstpage}\@thanks
 \endgroup
 \setcounter{footnote}{0}
 \let\maketitle\relax
 \let\@maketitle\relax
 \gdef\@thanks{}\gdef\@author{}\gdef\@title{}\let\thanks\relax}
\makeatother

\numberwithin{equation}{section}

\newcommand{\be}{\begin{equation}}
\newcommand{\bea}{\begin{eqnarray}}
\newcommand{\ee}{\end{equation}}
\newcommand{\eea}{\end{eqnarray}}

\newcommand{\Tr}{{\rm Tr}}
\newcommand{\tr}{{\rm tr}}
\renewcommand{\t}{\tilde}
\newcommand{\muphi}{\mu_\phi}

\begin{document}
 
\setcounter{page}0
\def\ppnumber{\vbox{\baselineskip14pt
}}
\def\ppdate{\footnotesize{NSF-KITP-11-077}} \date{}

\author{Carlos Tamarit\\
[7mm]
{\normalsize  \it Kavli Institute for Theoretical Physics}\\
{\normalsize \it University of California, Santa Barbara, CA 93106, USA}\\
[3mm]
{\tt \footnotesize tamarit at kitp.ucsb.edu}
}

\bigskip
\title{\bf Decays of metastable vacua in SQCD
\vskip 0.5cm}
\maketitle

\begin{abstract} \normalsize
\noindent The decay rates of metastable SQCD vacua in ISS-type models, both towards supersymmetric vacua as well as towards other nonsupersymmetric configurations arising in theories with elementary spectators, are estimated numerically in the semiclassical approximation by computing the corresponding multifield bounce configurations. The  scaling of the bounce action with respect to the most relevant dimensionless couplings and ratios of scales is analyzed. In the case of the decays towards the susy vacua generated by nonperturbative effects, the results confirm previous analytical estimations of this scaling, obtained by assuming a triangular potential barrier. The decay rates towards susy vacua generated by R-symmetry breaking interactions turn out to be more than sufficiently suppressed for the phenomenologically relevant parameter range, and their behavior in this regime differs from analytic estimations valid for parametrically small scale ratios. It is also shown that in models with spectator fields, even though the decays towards vacua involving nonzero spectator VEVs don't have a strong parametric dependence on the scale ratios, the ISS vacuum can still be made long-lived in the presence of R-symmetry breaking interactions.

\end{abstract}
\bigskip
\newpage


\section{Introduction}

Since Intriligator, Seiberg and Shih showed in their seminal paper \cite{Intriligator:2006dd} that long lived,  supersymmetry-breaking metastable vacua arose in simple theories, such as SQCD in the free magnetic phase, many works have studied models of particle physics based upon ISS-type constructions. A minimum requirement for these models to be considered as plausible descriptions of the particle interactions in Nature is that the lifetime of the relevant vacuum should be greater than the estimated age of the Universe. 

Ref.~\cite{Intriligator:2006dd} argued that the lifetime of the ISS vacuum  in the original model depended parametrically on the ratio between the dynamical scale $\Lambda$ of the gauge interactions and the mass scale $\mu$ of the low-energy magnetic theory, which is related to the masses of the electric quarks. Thus, long enough lifetimes are guaranteed for $\mu/\Lambda\ll1$, which can always be achieved by pushing $\Lambda$ to higher energies. A similar situation holds for models in which R-symmetry violating interactions involving a new scale $\mu_\phi$ are introduced, as in ref.~\cite{Essig:2008kz}; in this case the decay towards supersymmetric vacua is parametrically suppressed for $\mu_\phi/\mu\ll1$. However, since  $\muphi$ determines the gaugino masses while $\mu$ ends up controlling the squark soft masses, the ratio $\muphi/\mu$ cannot be made too small without creating a very fine-tuned spectrum. In principle it is not completely clear whether this will be in conflict with the vacuum being sufficiently stable, but the results of ref.~\cite{Essig:2008kz} showed that this is not necessarily the case. Here a more a more detailed numerical analysis will be presented.

Another threat to the stability of the vacuum appears when the ISS model is coupled to elementary spectators. In models based upon ISS constructions and attempting to link supersymmetry breaking with flavor physics, such as those of refs.~\cite{Franco:2009wf,Craig:2009hf,Behbahani:2010wh,SchaferNameki:2010iz}, some of the generations of the Standard Model are embedded into the meson fields of the low energy theory; generically, this cannot be done without having light mesons in undesired representations of the gauge group, which have to be lifted by coupling them to elementary spectators. These new couplings give rise to new metastable vacua involving nonzero expectation values for the spectator fields, which have lower energy than the ISS vacuum \footnote{These vacua were initially found by D. Green, A. Katz and Z. Komargodski. Spectator fields may also be added to avoid tachyonic squarks in dynamical realizations of minimal gauge mediation, as in ref.~\cite{Hook:2011ea}, though in this case they do not give rise to lower energy vacua. Other models use spectator fields to build dynamical models of gaugino mediation \cite{Green:2010ww}.} \cite{Behbahani:2010wh}. In contrast with the decays towards supersymmetric minima, which exhibit the parametric suppressions alluded to before, it is not clear whether the decay rates towards the new vacua will  have a strong dependence on the ratios of scales; at least for weak R-symmetry breaking interactions, the rates are expected to be
 mainly determined by a dimensionless coupling which cannot be made parametrically small in phenomenologically acceptable models. It was argued in refs.~\cite{Behbahani:2010wh,SchaferNameki:2010iz} that by making the electric quarks appropriately non-degenerate, the energy of the new vacua can be lifted above that of the ISS vacuum, which would prevent the decay. However, in models attempting to explain gaugino masses this solution is unsatisfactory, since the consideration of nondegenerate masses for the electric quarks leads to a suppression of the gaugino masses arising from R-symmetry breaking interactions \cite{SchaferNameki:2010iz}. Thus, in models with degenerate quark masses and spectator fields, with no extra mass scales involved in the interactions of the latter --otherwise it is straightforward to lift the vacua involving spectator VEVs above the ISS vacuum-- a more detailed estimation of the decay rates is needed in order to discern whether the ISS-type vacua are long-lived enough.

This paper presents new calculations of the decay rates of ISS vacua in models with degenerate electric quarks, both towards supersymmetric vacua and towards vacua involving expectation values for spectator fields, in theories with and without R-symmetry breaking interactions. These calculations were performed by solving numerically for the semiclassical bounce configurations that dominate the path integral representing the decay amplitude \cite{Coleman:1977py}. The computation is not straightforward due to the fact that the bounce configurations involve several fields; it was tackled following the approach of ref.~\cite{Konstandin:2006nd}, with up to five fields. Also, since some of the vacua are only stabilized after taking into account quantum corrections, these have to be taken into account when solving for the bounce configurations.  Partial results were already quoted in ref.~\cite{SchaferNameki:2010iz}; here a more complete presentation is given as well as results involving less approximations. 

The results confirm the qualitative dependence of the bounce action on dimensionless couplings and ratios of scales that can be extracted by identifying the small parameters that dominate the relevant contributions to the potential. Thus, the decays towards susy vacua generated by nonperturbative interactions are indeed parametrically suppressed for growing values of the ratio $\frac{\Lambda}{\mu}$. Similarly, the decays towards susy vacua generated by R-symmetry breaking interactions are suppressed for diminishing $\frac{\mu_\phi}{\mu}$. Finally, the decays towards vacua involving nonzero VEVs of spectator fields are controlled essentially by the dimensionless parameter $h$ present in the ISS model superpotential, although for fixed $h$ they are enhanced by increasing $\frac{\mu_\phi}{\mu}$. This last property could not be captured by the simplified calculations of ref.~\cite{SchaferNameki:2010iz}, which ignored R-symmetry breaking interactions, and used only a rough estimation of the gauge fields' one-loop contributions to the scalar potential. Analytic estimations of the scaling of the bounce action with respect to the above dimensionless parameters, that were previously obtained in the literature by assuming a triangular barrier and restricting to one field direction, are confirmed for the decays towards the nonperturbative susy vacua. In the case of susy vacua with $\muphi\neq0$, the analytical scaling obtained in ref.~\cite{Essig:2008kz} is valid for $\mu_\phi/\mu\ll1$, which is not the regime studied here --numerical fluctuations make it hard to explore. The scaling obtained here differs from the triangular barrier result, but the latter could still be valid for $\mu_\phi/\mu\ll1$. Regarding the resulting lifetimes, all decays to susy vacua, including the case with R-symmetry breaking interactions, turn out to be more than sufficiently suppressed for typical, phenomenologically acceptable values of the parameters, so that demanding long-lived vacua does not impose phenomenologically relevant constraints. In the case of models with spectator fields, centering on the case in which the magnetic group is trivial and for diagonal spectator fields (i.e. coupling to the diagonal of the meson field matrix), sufficiently long lifetimes can be achieved for small, nonzero $\muphi$ (perfectly safe regarding the decays towards susy vacua) and order one values of $h$; for the choice of parameters of \S\ref{subsec:numerics:spect}, the bounce action for sufficiently high values of $\frac{\muphi}{\mu}\gtrsim4\cdot10^{-3}$ scales roughly (up to errors around 10\%) as
\begin{align}\label{eq:spectscaling}
 S_b^{\rm ISS\rightarrow spect}\sim \frac{3\cdot 10^8\muphi^2}{h^{3.5}\mu^2}.
\end{align}
Regarding off-diagonal spectators (coupling to off-diagonal mesons), they are expected to generate milder instabilities than the diagonal ones: the new vacua are either further from the origin of field space or involve more fields getting a VEV, which implies an increase in the number of tunneling directions. Hence, if a model is safe with respect to decays towards vacua associated with diagonal spectator fields, it is expected that it will also be safe concerning decays to vacua generated by off-diagonal spectators. Thus only the first type decays were analyzed in detail.

Finally, it should be commented that there exist other potential sources of instabilities in ISS-type models, which do not involve tunneling processes but rather growing classical solutions, such as Q-balls. Ref.~\cite{Barnard:2010wk} argued that by making the dimensionless coupling $h$ of the ISS model greater than the gauge coupling $g$ of the gauged global symmetry, $h>g$, then Q-balls are unstable towards decay to massive vector bosons and they do not threat the stability of the theory. This requirement seems in tension with the fact that the decay rate towards spectator vacua grows for increasing values of h --see eq.~\ref{eq:spectscaling}. The above bound applies when the full global symmetry is gauged, so that the Goldstone modes of the global symmetry are eaten away and do not support Q-ball solutions. The situation worsens in models in which only a subgroup is gauged, since then the Goldstone modes become pseudo-Goldstone modes with masses of the order $g/4\pi$, which can support Q-balls and cannot decay to other particles. However, there is a possible way out consisting in gauging the global U(1) symmetry that guarantees the existence of Q-balls. This gauging seems to make the Q-balls unstable, at least for large values of the gauge coupling \cite{Coleman:1985ki,Benci:2010cs}.

The organization of the rest of the paper is as follows. Section \ref{sec:vacua} reviews the ISS model, including the ISS vacuum, the supersymmetric vacua and the vacua arising when spectator fields are added. Section \ref{sec:numerics} deals with the numerical calculations. \S \ref{subsec:numerics:gen} introduces the numerical method, while \S \ref{subsec:numerics:dimensionless} introduces a dimensionless parameterization of the bounce action, which allows to extract qualitative conclusions about the dependence of the latter with respect to the couplings and scale ratios of the theory. The results for decays towards susy vacua with and without R-symmetry breaking and for vacua involving spectator fields are given in  \S\S \ref{subsec:numerics:susyR}, \ref{subsec:numerics:susynoR} and \ref{subsec:numerics:spect}, respectively. An appendix has been included, providing formulae for the tree-level mass matrices and the Coleman-Weinberg potential.


\section{Vacuum configurations in ISS-type models \label{sec:vacua}}

This paper focuses on weakly coupled ISS-type supersymmetric models involving an $SU(\tilde N_c)$ gauge group, an $N_f\times N_f$ matrix $\Phi$ of singlet meson fields, and fields $q$ and $\t q$ --``magnetic quarks''-- in the fundamental and antifundamental representations of the gauge group. The superpotential for the basic model is \cite{Intriligator:2006dd}
\begin{equation}
\label{eq:Wmag}
 W_{\rm mag}= - h \tr (\hat \mu^2 \Phi) + h \tr (q \Phi \t q),
\end{equation}
where the suffix ``mag'' is chosen to emphasize the fact that these models arise as low energy descriptions of $SU(N)$ SQCD with $N_f$ massive flavors in the free magnetic phase, $N_c<N_f<\frac{3}{2}N_c$, with $\tilde N_c=N_f-N_c$ \cite{Seiberg:1994pq}. $\hat \mu$ is a matrix of mass parameters, which can be related to the mass matrix of the electric quarks in the dual theory as
\begin{align*}
 -h\hat\mu^2\sim \Lambda m,
\end{align*}
$\Lambda$ being the dynamical scale of the electric $SU(N)$ gauge group. For the rest of the paper, the focus will be on degenerate electric quarks, yielding $\hat\mu=\mu\,\mathbb{I}_{N_f}$.

Apart from the weakly coupled superpotential $W_{\rm mag}$, one has to take into account nonperturbative contributions generated by the gauge dynamics,
\begin{align}
\label{eq:Wnp}
 W_{np}=\tilde N_c(h^{N_f}\Lambda_m^{3\tilde N_c-N_f}\det \Phi)^{\frac{1}{\tilde N_c}},
\end{align}
where $\Lambda_m$ is the dynamical scale in the magnetic theory, related to $\Lambda$ as \cite{Intriligator:2006dd}
\begin{align*}
 \Big(\frac{\Lambda}{\Lambda_m}\Big)^{3 N_c-2N_f}\sim \frac{(-1)^{N_f-N_c}}{h^{N_f}}.
\end{align*}
The theory has a global $SU(N_f)\times SU(N_f)\times U(1)_V$ symmetry under which the fields transform as

\begin{center}
\begin{tabular}{c|ccc}
&$SU(N_f)$&$SU(N_f)$&$U(1)_V$\\
\hline
$q$ & $\Box$  & $1$ & $+1$\\
$\t q$ & $1$ & $\overline\Box$ &  $-1$\\
$\Phi$ & $\overline\Box$ & $\Box$ & $0$.
\end{tabular}
\end{center}
In typical models, as in applications to gauge mediation or models with composite generations, a diagonal subgroup $SU(N_k)_D$, $N_k\leq N_f$ is gauged, where $SU(N_k)$ contains the Standard Model group. Similarly, models usually involve the gauging of one or more $U(1)$ factors in order to lift massless Goldstone modes around the soon to be reviewed ISS vacuum \cite{Essig:2008kz}. 

  Furthermore, the possibility of coupling the meson field to a matrix $S$ of elementary spectator fields will be considered, as in the models of 
refs.~\cite{Franco:2009wf,Craig:2009hf,Behbahani:2010wh,SchaferNameki:2010iz}, without the introduction of further mass scales. This entails a superpotential contribution of the form
\begin{align}
\label{eq:Wspect}
W_{\rm spect}=\lambda \Lambda \Tr\,\Phi S.
\end{align}
To avoid spoiling the ISS vacuum, in which $\Phi$ acquires nonzero F-terms along the diagonal, $S$ is taken as a traceless matrix. If $q$ and $\t q$ are chosen to be in a reducible representation of $SU(N_k)_D$, $S$ can have components transforming in both vector representations of this group --along diagonal blocks-- and chiral ones --along off-diagonal blocks. 
  Finally, in models of supersymmetry breaking based upon ISS vacua and aiming to explain gaugino masses, it is necessary to introduce an explicit source of violation of R-symmetry. We will follow the approach of ref.~\cite{Essig:2008kz} and add the following contribution to the superpotential:
\begin{align}\label{eq:WR}
 W_R=\frac{1}{2}\,h^2\mu_\phi \Tr\, \Phi^2.
\end{align}
Throughout the rest of the paper it will assumed that the K\"ahler potential for the fields is canonical; this is justified since the models considered arise as weakly coupled infrared descriptions of the strongly coupled electric theories. 

\subsection{Vacua in the R-symmetric case \label{subsec:vacua:R}}

This section provides a brief review of the vacuum configurations of models without R-symmetry breaking, with superpotential of the form
\begin{align}
\label{eq:WtotnonR}
 W=W_{\rm mag}+W_{\rm np}+W_{\rm spect},
\end{align}
where the different terms are given in eqs.~\eqref{eq:Wmag}, \eqref{eq:Wnp} and \eqref{eq:Wspect}.

\subsubsection{Susy vacua}

The superpotential of eq.~\eqref{eq:WtotnonR} admits supersymmetric configurations, thanks to the nonperturbative contributions. These are \cite{Intriligator:2006dd}

\begin{align*}
 h\Phi=\mu \Big(\frac{\mu}{\Lambda_m}\Big)^{\frac{3\tilde N_c-N_f}{N_f-\tilde N_c}}
\end{align*}
with the rest of the fields having zero VEVs.

\subsubsection{ISS vacuum}\label{subsec:ISS}

Ignoring the nonperturbative contributions $W_{np}$ to the superpotential, which can be done when the fields take values near the origin, the ISS model breaks supersymmetry due to the rank condition:
\begin{equation}\label{eq:rankcond}
-F_{\Phi}^\dagger= - h \hat \mu^2\mathbb{I} + h \, q  \t q^\intercal \neq 0,
\end{equation}
because the two terms in the sum do not have the same rank; at most one cancel ${\rm rank}(\tilde q q)=\tilde N_c<N_f$ F-terms. Thus, supersymmetry is broken. Decomposing the meson and magnetic quarks in the following way, 
\begin{equation*}\label{eq:fielddecomp}
\Phi = \left( 
\begin{matrix}
Y_{\t N_c \times \t N_c} & Z^T_{\t N_c \times (N_f - \t N_c )} \\
\t Z_{(N_f - \t N_c) \times \t N_c} & X_{(N_f - \t N_c)  \times (N_f - \t N_c) } 
\end{matrix}
\right),
\end{equation*}
\begin{equation*}
\t q =\left(
\begin{matrix}
\t \chi_{\t N_c \times \t N_c} \\ \t \rho_{(N_f - \t N_c )  \times \t N_c} 
\end{matrix}\right)\;,\;\;
q =\left(
\begin{matrix}
\chi _{\t N_c \times \t N_c}\\ \rho_{(N_f - \t N_c )  \times \t N_c} 
\end{matrix}
\right),
\end{equation*}
then the resulting ISS vacuum is characterized by nonzero F-terms $F_X=-h\mu^2$, with 
\begin{equation}\label{eq:ISS}
\langle\t\chi\chi\rangle=\mu^2 \mathbb{I}_{\tilde N_c},
\end{equation}
and the rest of the fields having zero VEVs. The tree-level vacuum energy is
\begin{align*}
 V_{\rm ISS}=h^2 N_c\,\mu^4.
\end{align*}
To lowest order, $X$ is a pseudomodulus, but it is stabilized at the origin by one-loop quantum corrections \cite{Intriligator:2006dd}. It should be emphasized that all field directions, including the Goldstone modes associated with the global symmetries broken by the configuration of eq.~\eqref{eq:ISS}, are stabilized once one gauges the global U(1) symmetry of the model \cite{Essig:2008kz}. Thus, the decays to other vacua can only proceed through tunneling (as long as the Q-balls of ref.~\cite{Barnard:2010wk} are made unstable, as commented in the introduction).

\subsubsection{Vacua involving spectator VEVs}

As was motivated before, the spectator fields can be embedded into a traceless matrix. They are used to lift some mesons included in the field $X$ of eq.~\eqref{eq:fielddecomp}, which are light in the ISS vacuum; thus, $S$ only couples to $X$. Taking into account the spectator couplings, and ignoring again nonperturbative contributions to the superpotential, instead of eq.~\eqref{eq:rankcond} one has
\begin{equation}\label{eq:Ftermsspect}
-F_{Y}^\dagger= - h \hat \mu^2\mathbb{I}_{\t N_c} + h \, \chi  \t\chi^\intercal, \quad  -F_{X}^\dagger= - h \hat \mu^2\mathbb{I}_{N_c} + h \,\rho  \t \rho^\intercal+\lambda \Lambda S.
\end{equation}
The rank condition applies as before to the fields $\chi,\t\chi$ but not to $\rho,\t \rho$; one can use the spectators to cancel more F-terms than in the ISS vacuum.
 \cite{Behbahani:2010wh}. 

One can then consider VEVs for both off-diagonal as well as diagonal components of the spectator fields. In the case of diagonal VEVs, if $S$ is an $N_j\times N_j$ traceless matrix coupling to an $N_j\times N_j$ block $X_j$ of the meson field, one can use $N_j-1$ diagonal elements to cancel as many F-terms of $X_j$, and then the uncancelled F-term can be made zero by giving a VEV to a pair of components of the $\rho, \tilde\rho$ fields. This in turn requires to set one component of $\chi, \tilde\chi$ to zero in order not to turn on some components of $F_Z,\,F_{\tilde Z}$. The new vacuum is \cite{Behbahani:2010wh}
\begin{align}\label{eq:vacuumspect}
 \langle\tilde\chi\chi\rangle=\mu^2\mathbb{I}_{\t N_c-1},\quad \langle\t\rho\rho\rangle_{j\times j}=\mu^2{\rm diag}(0,\dots,0,j),\quad \langle S\rangle=\frac{\mu}{\lambda\Lambda}{\rm diag}(1,\dots,1,-(j-1)),
\end{align}
whose tree-level energy is 
\begin{align}\label{eq:vacumenergyspect}
 V_{\rm spect}=h^2(N_c-j+1)\,\mu^4,
\end{align}
i.e., lower than the energy of the ISS vacuum. In this vacuum, $Y$ as well as the components of $X$ that do not couple to spectators are tree-level flat directions; again, they are stabilized by quantum corrections. 

Regarding vacua with VEVs for off-diagonal spectators, one may distinguish two cases. The first case is that in which off-diagonal spectator fields come in conjugate pairs. This allows again to cancel more F-terms than the $\tilde N_c$ ones allowed by the rank condition for the ISS vacuum. Considering $\rho,\tilde\rho$ as $N_c\times\tilde N_c$ matrices, one can give identical VEVs to multiple rows of $\rho,\tilde\rho$; this allows to cancel the diagonal F-terms in eq.~\eqref{eq:Ftermsspect} but generates symmetric off-diagonal contributions, which can be compensated in turn by giving VEVs to off-diagonal spectators. The F-term conditions for  $F_Z,\,F_{\tilde Z}$ then force to set to zero one linear combination of the rows of the $\t N_c\times\tilde N_c$ matrix of VEVs of $\chi$ and $\t\chi$. Hence, if one cancels $j$ F-terms along $X$, one needs to give VEVs to $j$ rows of $\rho,\tilde\rho$ as well as $\frac{j(j-1)}{2}$ pairs of off-diagonal spectators; the resulting vacuum energy is again that of eq.~\eqref{eq:vacumenergyspect}.

Alternatively, one can have off-diagonal spectators that are not accompanied by their conjugates. Again, one can cancel the diagonal F-terms of $X$ as in the previous paragraph by giving VEVs to the $\rho,\t\rho$ fields, which generates the same pairs of off-diagonal contributions to  $F_X$. For each of these pairs, one can cancel one of the F-terms by giving a VEV to an off-diagonal spectator field; since it is assumed that the spectators do not come in conjugate pairs, there will remain a nonzero off-diagonal contribution, $-{F_{X}^\dagger}_{ij}=h \,\rho_i  \t \rho_j^\intercal\neq0,\,\,i\neq j$. As shown in \cite{SchaferNameki:2010iz}, the resulting vacuum configurations have a runaway towards vacua with energy as in eq.~\eqref{eq:vacumenergyspect}. To see this, consider the cancellation of a pair of diagonal components of  $-{F_{X}^\dagger}$, say $-{F_{X}^\dagger}_{ii}$ and $-{F_{X}^\dagger}_{jj}$, by choosing
\begin{align}\label{eq:runaway}
 \rho_i  \t \rho_i^\intercal=\rho_j  \t \rho_j^\intercal=\mu^2.
\end{align}
An off-diagonal spectator $S_{ji}$ can be given a VEV to cancel $(F_{X}^\dagger)_{ji}$; then $(F_{X}^\dagger)_{ij}=h \,\rho_i  \t \rho_j^\intercal$ is in principle nonzero but can be minimized by taking either $\rho_i$ or $\t\rho_j$ to zero, which by virtue of \eqref{eq:runaway} implies that either $\t\rho_i$ or $\rho_j$ must approach infinity, which generates a runaway.

At this point it should be clear that the tunneling from the ISS vacuum towards the vacua of this section should proceed preferably towards those involving spectators along the diagonal, either because they are closer to the ISS vacuum or because the decay involves a smaller number of tunneling directions. For this reason the numerical estimates in \S\ref{sec:numerics} center on the case of diagonal spectators.

\subsection{Vacua in the presence of R-symmetry violating interactions \label{subsec:vacua:nonR}}

This section is dedicated to the vacuum configurations in the presence of R-symmetry breaking interactions coming from the superpotential contribution of eq.~\eqref{eq:WR}, so that the total superpotential is of the form
\begin{align*}
 W=W_{\rm mag}+W_{\rm np}+W_{\rm spect}+W_{\rm R},
\end{align*}
where the different terms are given in eqs.~\eqref{eq:Wmag}, \eqref{eq:Wnp}, \eqref{eq:Wspect} and \eqref{eq:WR}.

In this case, the violation of R-symmetry gives rise to susy-vacua near the origin, as expected from the results of ref.~\cite{Nelson:1993nf}. These vacua are given by
\begin{align*}
 \Phi=\frac{\mu}{h\mu_\phi}.
\end{align*}

If the R-symmetry breaking deformation $W_{\rm R}$ of the superpotential is small, it is expected that there will be vacuum configurations deforming the ISS vacuum of \S \ref{subsec:ISS}. This turns out to be the case indeed; although at tree-level the pseudomodulus field $X$ of the ISS model becomes unstable, one-loop quantum corrections stabilize it at a nonzero VEV, so that the new vacuum becomes \cite{Essig:2008kz}
\begin{align}\label{eq:vevsR}
 \langle \t\chi\chi\rangle=\mu^2\mathbb{I}_{\t N_c},\quad X\sim \frac{16\pi^2\mu_\phi}{h \tilde N_c}.
\end{align}

Analogously, the vacua involving spectator fields also survive the $\mu_\phi$ deformation; in this case the pseudomodulus field $Y$ gets a nonzero VEV due to quantum corrections, of the same order as $X$ in eq.~\eqref{eq:vevsR}. More details about the Coleman-Weinberg potential responsible for these quantum corrections are given in \S \ref{app:VCW}.


\section{Numerical estimations of decay rates \label{sec:numerics}}
\subsection{Method \label{subsec:numerics:gen}}

In order to estimate the decay rates of the ISS vacuum towards the supersymmetric and nonsupersymmetric vacua describe above, the semiclassical approach of Coleman \cite{Coleman:1977py} will be followed. In this approach, the lifetime of a metastable vacuum is computed from a saddlepoint evaluation of the path integral for the transition amplitude of the false vacuum onto itself after infinite time. In a theory involving a set of fields $\phi=\{\phi_i\}$, the decay rate per unite volume can be written as \cite{Coleman:1977py,Callan:1977pt}
\begin{align}\label{eq:Gamma}
 \frac{\Gamma}{V}=e^{-\frac{S_b}{\hbar}}\frac{S_b^2}{4\pi^2\hbar^2}\left|\frac{\det'[-\Box+V''(\overline\phi)]}{\det[-\Box+V''(\phi_-)]}\right|^{-1/2}\times (1+O(\hbar)), \,\,\,\,S_b=S(\overline\phi)-S(\phi_-).
\end{align}
In the previous formula, $\phi_-$ denotes the metastable vacuum configuration, while $\det'$ indicates the determinant with translational zero modes omitted. $\overline\phi$ represents the semiclassical bounce configuration, which is obtained as a solution of the Euclidean equations of motion with appropriate boundary conditions. With the Euclidean action written as
\begin{align}\label{eq:Euclidact}
 S[\phi_i]=\int\! d^4x \frac{1}{2}\sum_i(\partial_\mu \phi_i)^2+V(\phi_i),
\end{align}
then the bounce configuration satisfies the equations
\begin{align*}
 \Box \,\phi_i-\frac{\partial V(\phi)}{\partial{\phi_i}}=0,\quad
\lim_{\tau\rightarrow\infty}\phi(\tau,\vec{x})=\phi_-,\quad\partial_\tau\phi(0,\vec{x})=0.
\end{align*}
Assuming an $O(4)$ symmetry, writing $r=(\tau,\vec{x})$, the equations above turn into
\begin{align}\label{eq:bounceq}
 \frac{d^2\phi_i}{dr^2}+\frac{3}{r}\frac{d\phi_i}{dr}=\frac{\partial V(\phi)}{\partial\phi_i},\quad \lim_{r\rightarrow\infty}\phi=\phi_-,\quad \frac{d\phi}{dr}(0)=0.
\end{align}
In the case of a single field, eq.~\eqref{eq:bounceq} can be solved numerically by employing a shooting method. However, this is not suited for multifield configurations. The calculations presented in this paper were done with the technique of ref.~\cite{Konstandin:2006nd}. Essentially, this approach contemplates the equation \eqref{eq:bounceq} with a modified friction term, 
\begin{align}\label{eq:bouncealpha}
 \frac{d^2\phi_i}{dr^2}+\frac{\alpha-1}{r}\frac{d\phi_i}{dr}=\frac{\partial V(\phi)}{\partial\phi_i}.
\end{align}
The equations for the frictionless case $\alpha=1$ are equivalent to the equations of motion of the following one-dimensional action
\begin{align}\label{eq:Stilde}
 \tilde S=\int_0^\infty\! dr\,\Big[\frac{1}{2}\sum_i\Big(\frac{d\phi_i}{dr}\Big)^2+V(\phi)\Big].
\end{align}
However, the bounce solution is not strictly a minimum of the above functional, but rather corresponds to a saddlepoint \cite{Callan:1977pt} --solutions staying closer to the true vacuum than the bounce have lower values of  $\tilde S$. Nevertheless, by using a modified potential $V_\epsilon$, obtained from $V$ by flattening out the potential energies below that of the false vacuum, it can be argued that the frictionless bounce configuration for $\alpha=1$ (see eq.~\eqref{eq:bouncealpha}) is a minimum of the functional $\tilde S_\epsilon$ obtained by substituting $V$ with $V_\epsilon$ in eq.~\eqref{eq:Stilde}, in the limit $\epsilon\rightarrow0$ \cite{Konstandin:2006nd}. $V_\epsilon$ can be taken as
\begin{align*}
 V_\epsilon=\frac{V(\phi)-V(\phi_-)}{2}+\sqrt{\frac{(V(\phi)-V(\phi_-))^2}{4}+\epsilon^2}-2\epsilon\frac{|\phi-\phi_-|^3}{|\phi_+-\phi_-|^3}+3\epsilon\frac{|\phi-\phi_-|^2}{|\phi_+-\phi_-|^2},
\end{align*}
where $\phi_+$ designates the lower lying vacuum. The reason for this is that, for  $\epsilon\rightarrow0$, the frictionless bounce has the same value of $\tilde S_\epsilon$ than field configurations joining both vacua and hence minimizing the functional; the bounce can be obtained from them by simply eliminating the points with $V(\phi)<V(\phi_-)$, whose contribution to the action vanishes for  $\epsilon\rightarrow0$.

Once the solution of eq.~\eqref{eq:bouncealpha} for $\alpha=1$ has been obtained via a minimization procedure, the sought for solution corresponding to $\alpha=4$ can be obtained iteratively in incremental steps in $\alpha$ by solving a finite-difference version of eq.~\eqref{eq:bounceq} with a linearized gradient of the potential. Discretizing the $r$ variable as an $n$ point lattice $r_k= k \Delta r, k=0,\dots,n-1,$ and denoting $\phi_k(r_i)\equiv\phi^i_k$, the linearized equation for a given value of $\alpha$ and a previous estimation $\tilde\phi$ of the solution is
\begin{align*}
 \frac{\phi^{i+1}_k-2\phi_k^{i}+\phi^{i-1}_k}{\Delta r^2}+\frac{\alpha-1}{(i+\Delta i)\Delta r}\frac{\phi^{i+1}_k-\phi^{i-1}_k}{2\Delta  r}=\partial_k V(\t\phi^i)+(\phi^i_l-\t\phi^i_l)\partial_l\partial_k V(\t\phi^i),
\end{align*}
which can be easily inverted. $\Delta i$ is an offset parameter introduced to avoid pathological behavior near $i=0$, which eventually has to be driven to zero. In the computations presented here, the iterations were performed with increasing number of steps in $\alpha$ until the successive values of the bounce action differed in less than a few percent.

In the previous formulae,  $V(\phi)$ is in principle meant to be the tree-level potential. However, when the higher energy vacuum is only stabilized after taking into account quantum corrections --as happens in the $\mu_\phi$ deformation of the ISS vacuum-- there will be no bounce solution to eq.~\eqref{eq:bounceq}. An obvious way out is to substitute the tree-level potential by the perturbative quantum effective potential $V_{\rm eff}(\phi)$ \cite{Frampton:1976kf}. This approach is correct as long as the effect of higher derivatives terms or field renormalizations can be ignored; to take these into account, and to avoid problems with imaginary values of the effective potential, a modified formalism is needed, such as the one described in ref.~\cite{Weinberg:1994tk}. To leading order both approaches are expected to be equivalent; therefore in the computations presented here it was proceeded by simply using the one-loop effective potential, which was obtained by evaluating the Coleman-Weinberg potential for arbitrary background values of the fields, as detailed in \S\ref{app:VCW}.

The numerical method summarized above allows to compute the bounce action $S(\overline\phi)$ in eq.~\eqref{eq:Gamma}. The calculation of the prefactor is a much more daunting task; however, since the bounce action enters exponentially, its contribution is dominant. The prefactor can be estimated as an appropriate power of the most relevant scale in the potential; with scales of the order of 200 TeV --the typical mass scale for the fields $\rho,Z$ in phenomenological models of direct mediation based on the ISS model, then a bounce action of the order $S_b\equiv S(\overline\phi)-S(\phi_-)\gtrsim400$ (in natural units) is enough to guarantee a lifetime of the order of the age of the Universe.

\subsection{Dimensionless parametrization of the bounce action\label{subsec:numerics:dimensionless}}

Since the action of the bounce configuration is dimensionless in natural units, it depends only on dimensionless couplings or ratios of the scales in the potential. Thus, it is desirable to perform the computations using dimensionless variables. This not only might help avoid numerical fluctuations in the presence of very different mass scales, but also allows to reach qualitative conclusions about the dependence of the bounce action on the dimensionless quantities.

The parameters in the superpotentials of eqs.~\eqref{eq:Wmag}, \eqref{eq:Wnp}, \eqref{eq:Wspect} and \eqref{eq:WR} are the dimensionless couplings $h,\lambda$ and the scales $\mu,\muphi,\Lambda$. From these we can define dimensionless scale ratios and fields, as well as a dimensionless radial variable and potential, as follows
\begin{align*}
 \kappa_\phi=\frac{\muphi}{\mu},\,\kappa_\Lambda=\frac{\Lambda}{\mu}, \quad \hat\phi=\frac{1}{\mu}\phi, \quad \hat r=\mu r, \quad\hat V(\hat\phi;h,\lambda,\kappa_\phi,\kappa_\lambda)=\frac{1}{\mu^4}V(\phi;h,\lambda,\muphi,\Lambda).
\end{align*}
Then the ``hatted'' fields corresponding to the bounce configuration satisfy the following equation involving only dimensionless quantities,
\begin{align*}
 \frac{d^2\hat\phi_i}{d\hat r^2}+\frac{3}{\hat r}\frac{d\hat\phi_i}{d\hat r}=\frac{\partial \hat V(\hat\phi)}{\partial\hat\phi_i},\quad \lim_{\hat r\rightarrow\infty}\hat\phi=\hat\phi_-,\quad \frac{d\hat\phi}{d\hat r}(0)=0,
\end{align*}
whereas the bounce action can be written as
\begin{align*}
 S=2\pi^2\int_0^\infty\! \hat r^3 d\hat r\,\Big[\frac{1}{2}\sum_i\Big(\frac{d\hat\phi_i}{d\hat r}\Big)^2+\hat V(\hat\phi)\Big].
\end{align*}
One can get some qualitative insight on the dependence of the bounce action on some of the dimensionless parameters when these get small \cite{Weinberg:1994tk}. Suppose that one can single out a coupling $\delta$ such that $\t V\equiv\frac{1}{\delta}\hat V$ is of order one. Rescaling the radial variable $\tilde r=\sqrt{\delta} \hat r$, the bounce action is
\begin{align*}
 S=\frac{2\pi^2}{\delta}\int_0^\infty\! \t r^3 d\t r\,\Big[\frac{1}{2}\sum_i\Big(\frac{d\hat\phi_i}{d\t r}\Big)^2+\t V(\hat\phi)\Big],
\end{align*}
where the integrand is expected to be of order one, so that the bounce action will scale like $\frac{1}{\delta}$.

In the case of decays from the ISS vacuum, generically $h$ is assumed to be of order one; since it appears in the F-terms and the masses of the messenger fields, it cannot be made too small in phenomenologically acceptable models. The parameters which can be made small are $\kappa_\phi$ and $\frac{1}{\kappa_\Lambda}$, which are expected to determine the decay rates towards the susy vacua. This is so because both parameters appear in the superpotential contributions $W_{np}$ and $W_R$ of eqs.~\eqref{eq:Wnp} and \eqref{eq:WR}, which are responsible for the existence of the susy vacua of \S\ref{subsec:vacua:R} and \S\ref{subsec:vacua:nonR}. These superpotentials give rise to terms  with positive powers of  $\frac{1}{\kappa_\Lambda}$ and $\kappa_{\muphi}$  in the potential, and the discussion in the previous paragraph applies. Thus, the bounce action is expected to be parametrically large for small values of these ratios of scales, which guarantees a parametrically small decay rate. The same conclusions were achieved in refs.~\cite{Intriligator:2006dd} and \cite{Essig:2008kz} using more detailed arguments based on one-dimensional tunneling with a triangular potential barrier, which yielded analytic estimates for the scaling. Ref.~\cite{Intriligator:2006dd} argued that the bounce action associated with the decay towards the nonperturbative susy vacua should behave as
\begin{align}\label{eq:scalingsusynp}
 S_b\sim \kappa_\Lambda^{\frac{4(N_f-3\t N_c)}{N_f-\t N_c}},
\end{align}
while in ref.~\cite{Essig:2008kz} the bounce action corresponding to the decay towards the susy vacua in the presence of R-symmetry breaking was estimated, in the limit $\kappa_\phi\ll1$ as
\begin{align}\label{eq:scalingsusyR}
 S_b\sim \frac{1}{\kappa_\phi^4}.
\end{align}
The next section will present numerical results which confirm the qualitative behavior and the scaling of eq.~\eqref{eq:scalingsusynp}. Regarding the decays to susy vacua in the presence of R-symmetry breaking, numerical fluctuations prevented to probe the regime $\kappa_\phi\ll1$ for which eq.~\eqref{eq:scalingsusyR} is valid. Still, a phenomenologically relevant range of parameters could be studied; the obtained scaling is different from eq.~\eqref{eq:scalingsusyR}, but it is still an open question whether the latter is valid or receives corrections due to the failure of the one-dimensional triangular barrier approximation.

Regarding the decays from the ISS vacuum to the vacua involving spectator fields, the situation seems less clear. From the scale ratios, only $\kappa_\phi$ will be relevant, since it influences the tunneling along the fields $X$ and $Y$, whose VEVs are controlled by $\muphi$. However, it is hard to guess how dominant the $\muphi$-dependent pieces in the potential are in comparison with other terms, and if at all, the decay rate is expected to increase for growing $\muphi$, since the difference between the VEVs in the ISS and spectator vacua grows with $\muphi$.  Concerning  $\lambda$ and $\kappa_\Lambda$, although they control the interaction with the spectator fields, typically one has $\kappa_\Lambda\gg1$, which  implies that the VEVs of the spectator fields will be very small --see eq.~\eqref{eq:vacuumspect}; thus the contribution of their kinetic energy to the bounce action will be subdominant, and their profile in the bounce configuration will be approximately that which minimizes the potential for the rest of the fields at any given point in the bounce trajectory. The value of the potential at these minima won't be affected by changes of $\lambda,\kappa_\Lambda$, and so they will have a weak influence in the final value for the bounce action. Thus, one expects that the latter will be mainly determined by the remaining dimensionless parameter $h$, which as said before cannot be made parametrically small for phenomenological reasons. Therefore, a more detailed numerical evaluation is needed in order to elucidate whether sufficiently long lifetimes can be achieved for phenomenologically acceptable values of the parameters.

The following sections present the results of the numerical evaluations of the bounce configurations and their actions. Given the preceding discussion, 
the computations are done for different values of $\kappa_\Lambda,\muphi$ and $h$, keeping fixed $N_c=5$ as well as $\tilde N_c=1$ and $\lambda=1$.

\subsection{Numerical results: decays towards susy vacua with no R-symmetry breaking \label{subsec:numerics:susynoR}}

\begin{figure}[h]\centering
\includegraphics[scale=1]{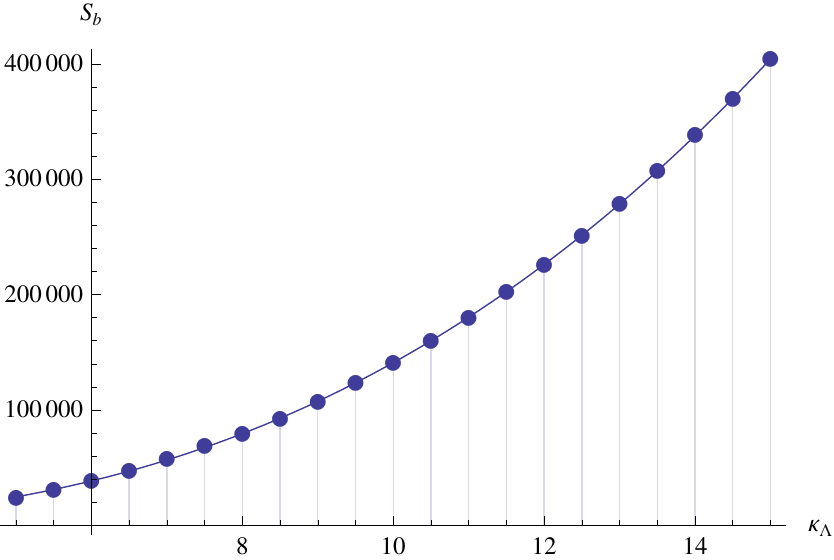}
\caption{\label{fig:1}Bounce action corresponding to the decay towards the susy vacua generated by nonperturbative interactions, for different values of $\kappa_\lambda$, $N_c=5,\t N_c=1,h=1,\kappa_\phi=0$. The solid line represents the function $S_b=332 \kappa_\Lambda^{2.6}+2068$.}
\end{figure}

\begin{figure}[h]\centering
\begin{minipage}{0.49\textwidth}\centering
 \includegraphics[scale=0.8]{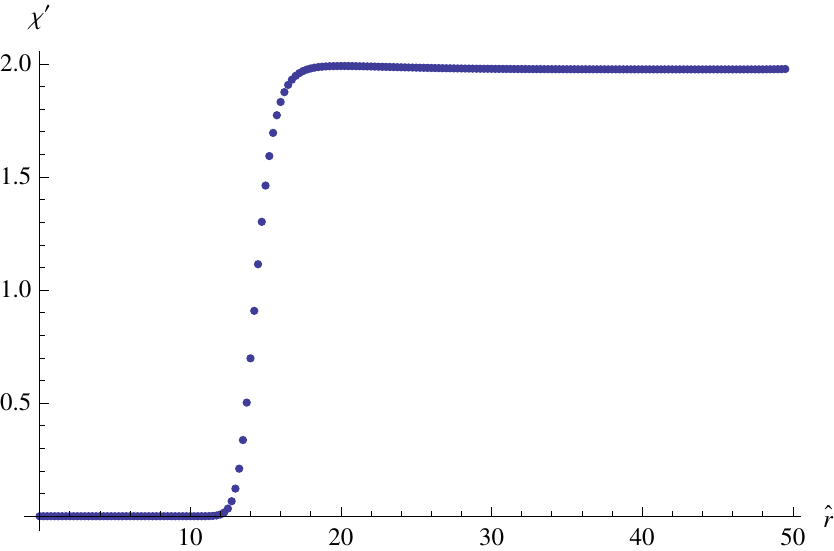}
\end{minipage}%
\begin{minipage}{0.49\textwidth}\centering
 \includegraphics[scale=0.8]{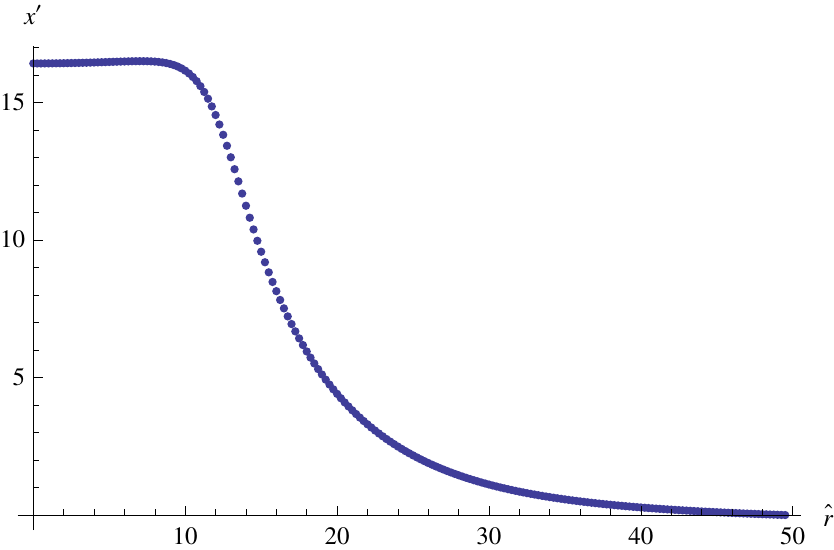}
\end{minipage}
\begin{minipage}{0.49\textwidth}\centering
 \includegraphics[scale=0.8]{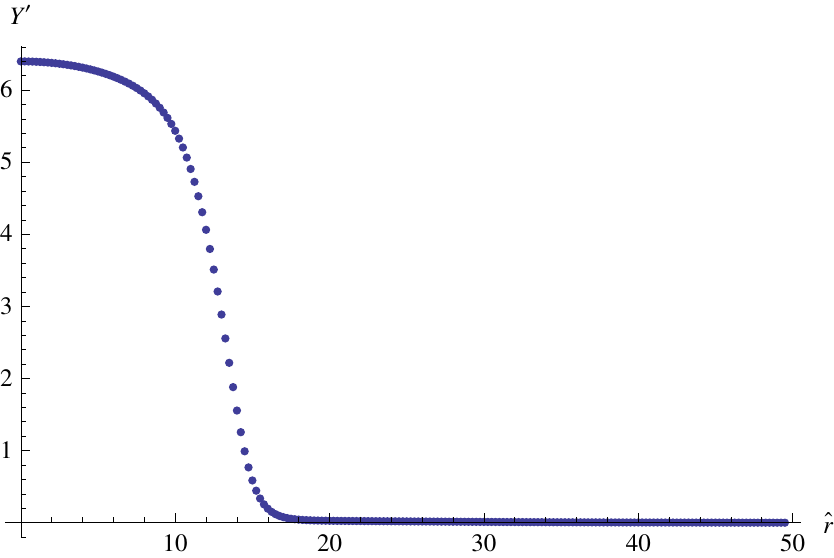}
\end{minipage}
\begin{minipage}{0.49\textwidth}\centering
 \includegraphics[scale=0.8]{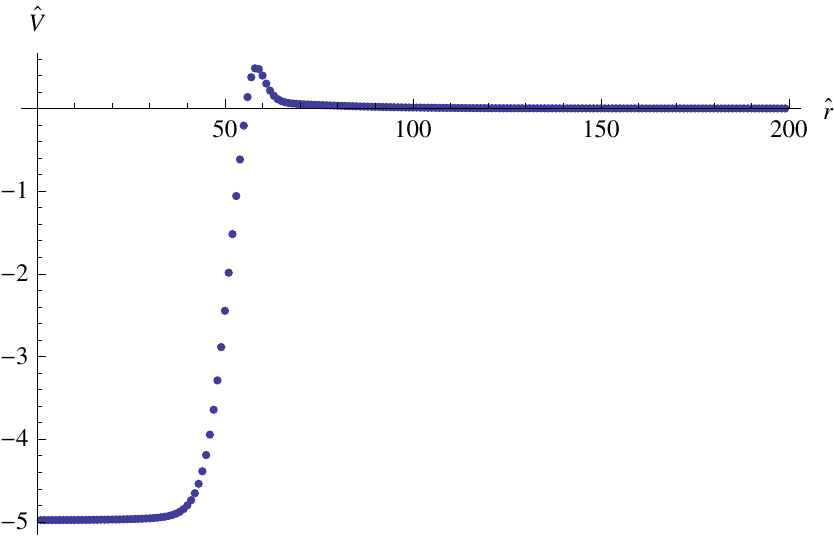}
\end{minipage}
\caption{\label{fig:2}Bounce configuration and potential corresponding to the decay towards the susy vacua generated by nonperturbative interactions, for  $N_c=5,\t N_c=1,h=1,\kappa_\phi=0, \kappa_\lambda=15$.}
\end{figure}

This section presents the results for the numerical evaluation of the bounce action associated with transitions from the ISS to the susy vacua generated by nonperturbative interactions, in the case $\mu_\phi=0$. Both vacua differ in the VEVs of the fields $X$, $Y$, $\chi,\t\chi$; in both of them one can choose $\langle\chi\rangle=\langle\t\chi\rangle$, which will be respected by the evolution under the equations of motion of the bounce. Turning on other fields apart from the above will increase the potential energy by turning on additional F-terms; thus the bounce configuration will only involve nontrivial profiles for three fields: $X$, $Y$ and $\chi=\t\chi$. The parameters in the superpotential are taken as real, which allows to only consider tunneling along the real  parts of the above fields; furthermore, $X$ is treated as a diagonal field, $X=x\mathbb{I}_5$. In order to solve for the bounce configuration with the technique of \S\ref{subsec:numerics:gen}, the fields have to be rescaled so that their kinetic terms are canonical as in eq.~\eqref{eq:Euclidact}; thus, the following redefinitions are introduced,
\begin{align}\label{eq:rescalesusy}
 x=\frac{1}{\sqrt{10}}\,x', \quad Y=\frac{1}{2}\,Y',\quad \chi=\t\chi=\frac{1}{2}\,\chi'.
\end{align}
The bounce action was computed for different values of the parameter $\kappa_\Lambda$, which according to the discussion in the previous section should largely determine the decay rate. The result is shown in Fig.~\ref{fig:1}, which confirms the parametric dependence in $\kappa_\Lambda$, with exponent equal to 2.6. This confirms the analytical result of eq.~\eqref{eq:scalingsusynp}, which for the choice of parameters here yields an exponent of 2.4. The results show that the normalization is such that even for moderate values of $\kappa_\Lambda$ the lifetimes are much bigger than the age of the Universe, $S_b\gg400$. One of the resulting multifield bounce configurations is shown on
 Fig.~\ref{fig:2}. 

\subsection{Numerical results: decays towards susy vacua with R-symmetry breaking \label{subsec:numerics:susyR}}

\begin{figure}[h]\centering
\includegraphics[scale=1]{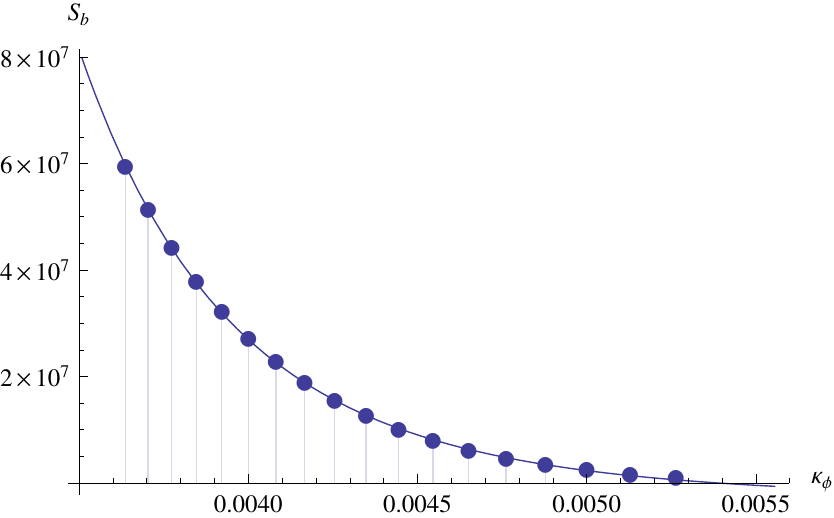}
\caption{\label{fig:3}Bounce action corresponding to the decay towards the susy vacua generated by R-symmetry breaking interactions, for different values of $\kappa_\phi$, $N_c=5,\t N_c=1,h=1,\kappa_\Lambda=3.85\cdot10^5$. The solid line represents the function $S_b=7.25\cdot10^{-12}{\kappa_\phi^{-7.76}}-2.98\cdot10^6$.}
\end{figure}

\begin{figure}[h]\centering
\begin{minipage}{0.49\textwidth}\centering
 \includegraphics[scale=0.8]{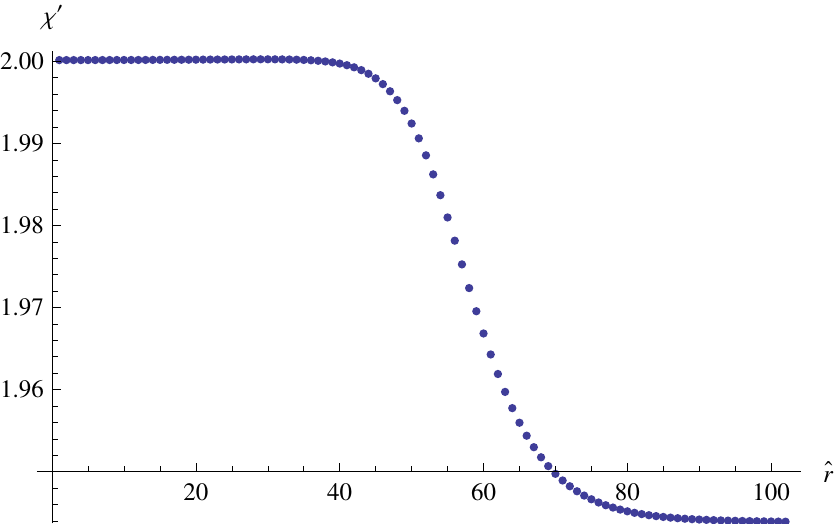}
\end{minipage}%
\begin{minipage}{0.49\textwidth}\centering
 \includegraphics[scale=0.8]{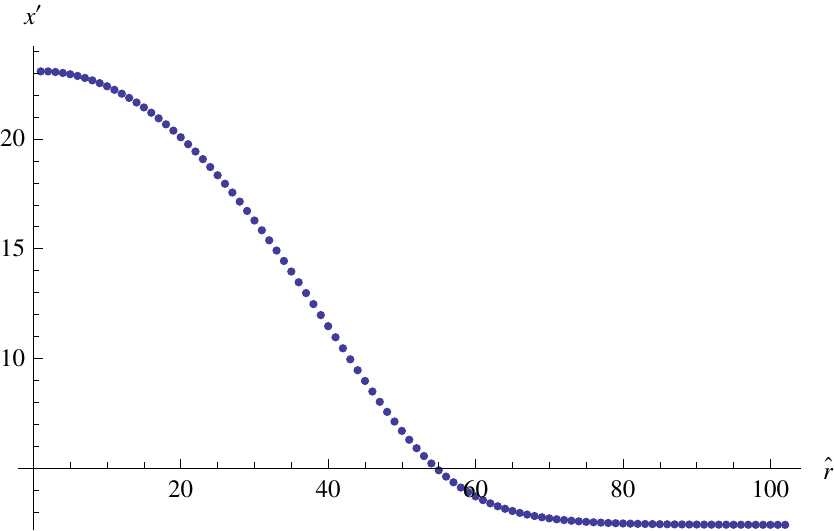}
\end{minipage}
\begin{minipage}{0.49\textwidth}\centering
 \includegraphics[scale=0.8]{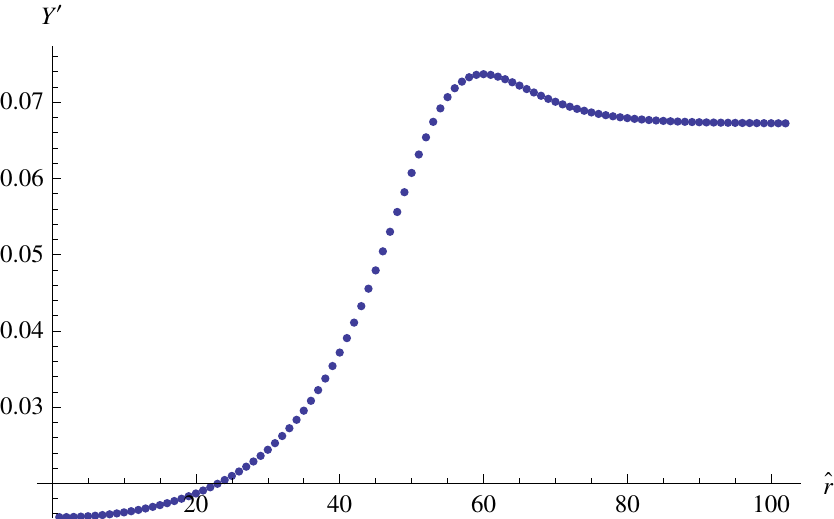}
\end{minipage}
\begin{minipage}{0.49\textwidth}\centering
 \includegraphics[scale=0.8]{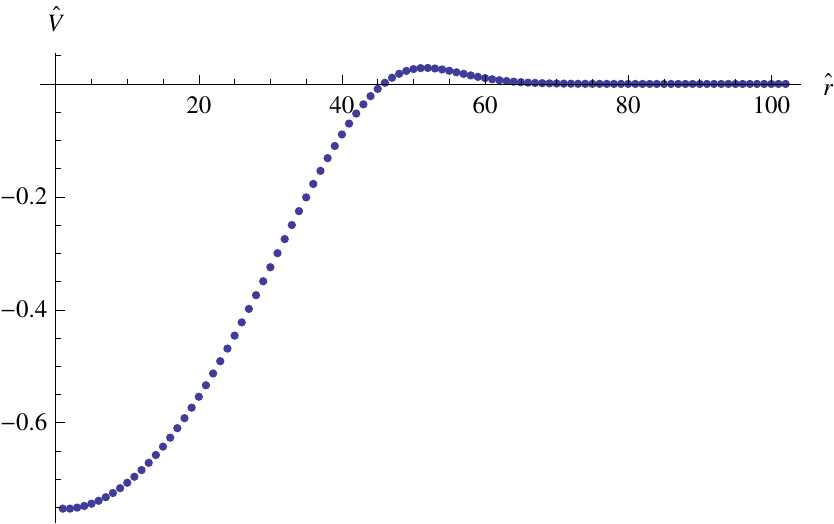}
\end{minipage}
\caption{\label{fig:4}Bounce configuration and potential corresponding to the decay towards the susy vacua generated by R-symmetry breaking interactions, for  $N_c=5,\t N_c=1,h=1,\kappa_\phi=0.005, \kappa_\lambda=3.85\cdot10^5$,.}
\end{figure}

By similar arguments to those employed in the previous section, one can conclude that the bounce configuration will again include only nontrivial profiles for the fields $\chi=\t\chi,X,Y$. Again, the rescalings of eq.~\eqref{eq:rescalesusy} are used, and the bounce action is computed for different values of the parameter $\kappa_\phi$, keeping $N_c=5,\t N_c=1,h=1$, as well as choosing a very high value of $\kappa_\Lambda=3.85\cdot10^5$ in order to make the nonperturbative interactions irrelevant. The results are shown in Fig.~\ref{fig:3}, which shows that the bounce action increases very rapidly for small $\kappa_\phi$, following a power law with exponent approximately equal to 8, which differs from the estimate of eq.~\eqref{eq:scalingsusyR}. This estimate is only valid for $\kappa_\phi\ll1$, which could not be probed due to numerical fluctuations; hence the discrepancy should not be too surprising. An example of a bounce configuration is shown on Fig.~\ref{fig:4}. Though the scaling with $\kappa_\phi$ differs from that of eq.~\eqref{eq:scalingsusyR}, qualitatively the bounce configuration behaves as described in ref.~\cite{Essig:2008kz}: $\chi'$ and $Y'$ only have small variations and the tunneling proceeds mainly in the $x'$ direction. It should be noted that the enhanced growth of the bounce action with diminishing $\kappa_\phi$ means that the lifetime of the ISS vacuum  will be long enough for typical phenomenologically acceptable models. In these, $\mu_\phi$ fixes the gaugino masses while $\mu$ determines the masses of the messenger fields, which gives $\mu_\phi\sim1{\rm TeV}, \mu\sim200{\rm TeV}$ and hence $\kappa_\phi\sim 0.005$, which is the region studied here, which yields $S_b\gg400.$ Thus one expects no phenomenologically relevant constraints coming from demanding sufficiently suppressed decays.

\subsection{Numerical results: decays towards vacua with nonzero VEVs for spectator fields \label{subsec:numerics:spect}}

\begin{figure}[h]\centering
\includegraphics[scale=1]{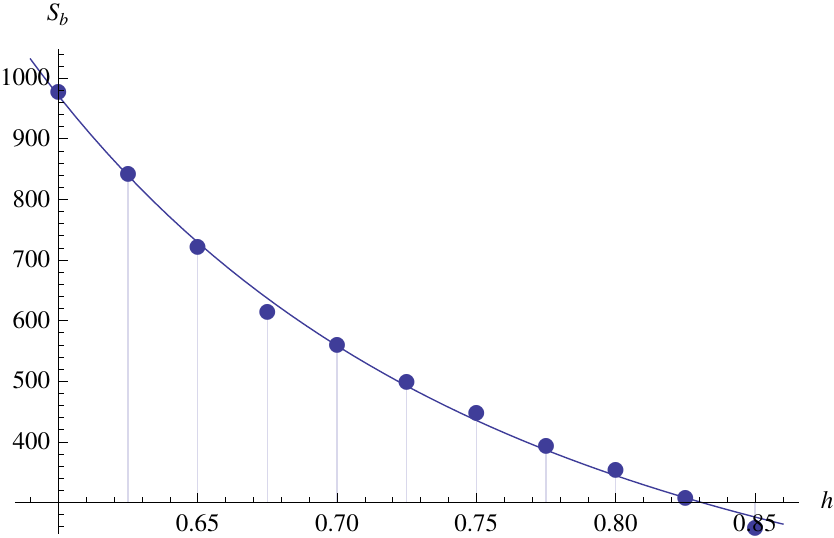}
\caption{\label{fig:5}Bounce action corresponding to the decay towards the vacua with nonzero spectator VEVs, for different values of $h$, $N_c=5,\t N_c=1,\kappa_\Lambda=38.5, \kappa_\phi=7.7\cdot10^{-4}$. The solid line represents the function $160.728 h^{-3.54115}-9.79449$.}
\end{figure}

\begin{figure}[h]\centering
\includegraphics[scale=1]{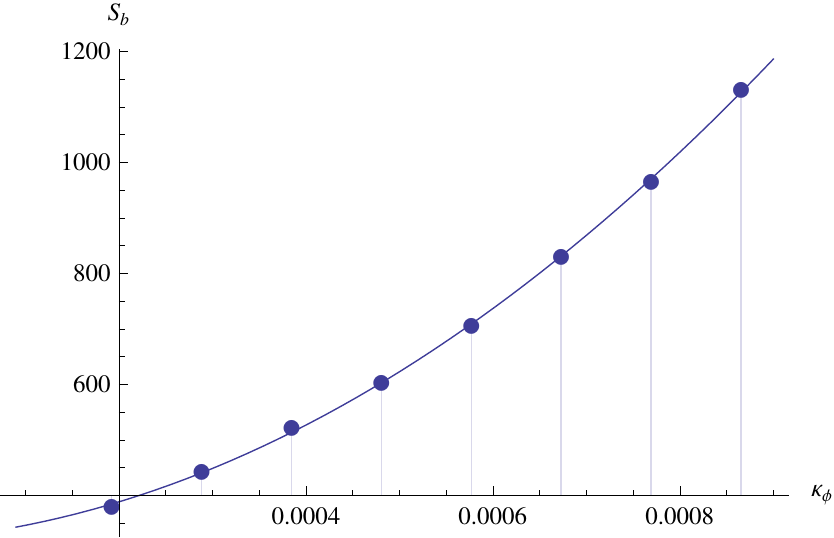}
\caption{\label{fig:6}Bounce action corresponding to the decay towards the vacua with nonzero spectator VEVs, for different values of $\kappa_\phi$, $N_c=5,\t N_c=1,h=0.6,\kappa_\Lambda=38.5$. The solid line represents the function $8.9719\times 10^8 \kappa_\phi^2+153663. \kappa_\phi+321.42$.}
\end{figure}

\begin{figure}[!h]\centering
\begin{minipage}{0.49\textwidth}\centering
 \includegraphics[scale=0.9]{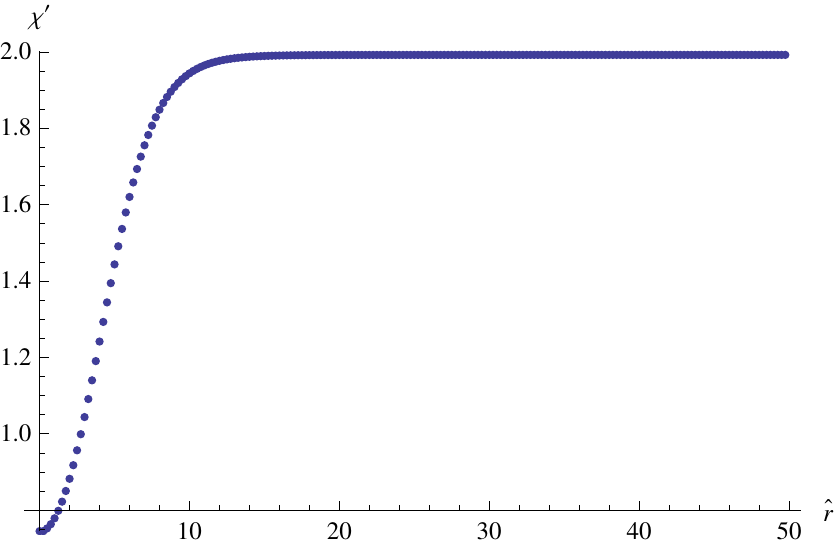}
\end{minipage}%
\begin{minipage}{0.49\textwidth}\centering
 \includegraphics[scale=0.9]{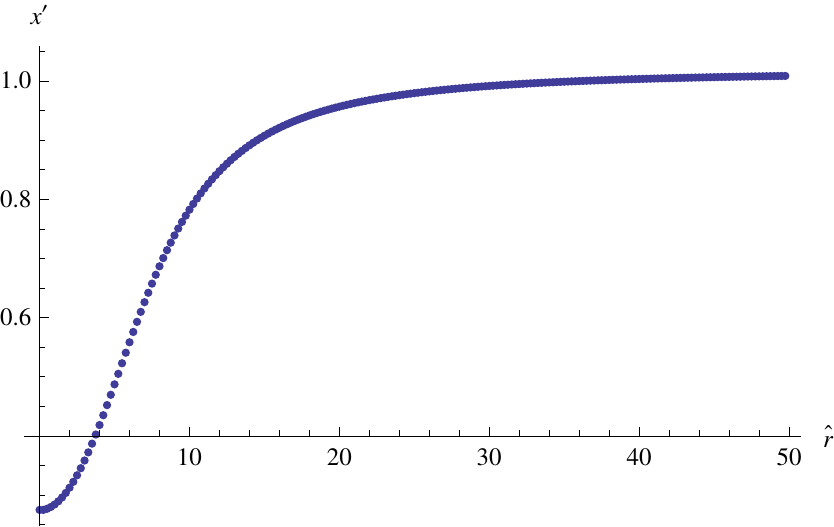}
\end{minipage}
\begin{minipage}{0.49\textwidth}\centering
 \includegraphics[scale=0.9]{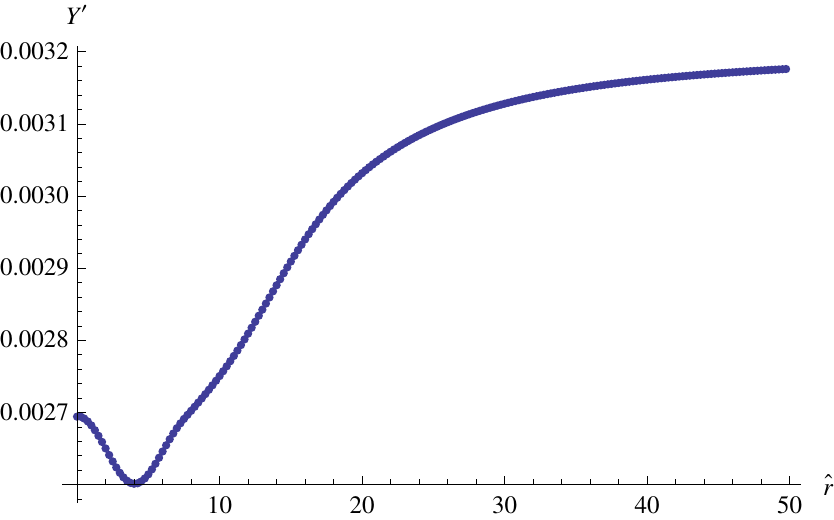}
\end{minipage}
\begin{minipage}{0.49\textwidth}\centering
 \includegraphics[scale=0.9]{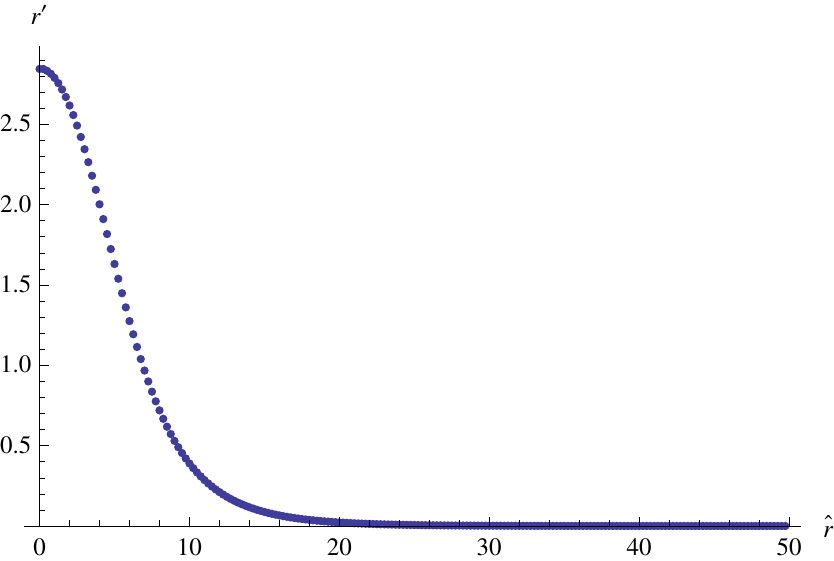}
\end{minipage}
\begin{minipage}{0.49\textwidth}\centering
 \includegraphics[scale=0.9]{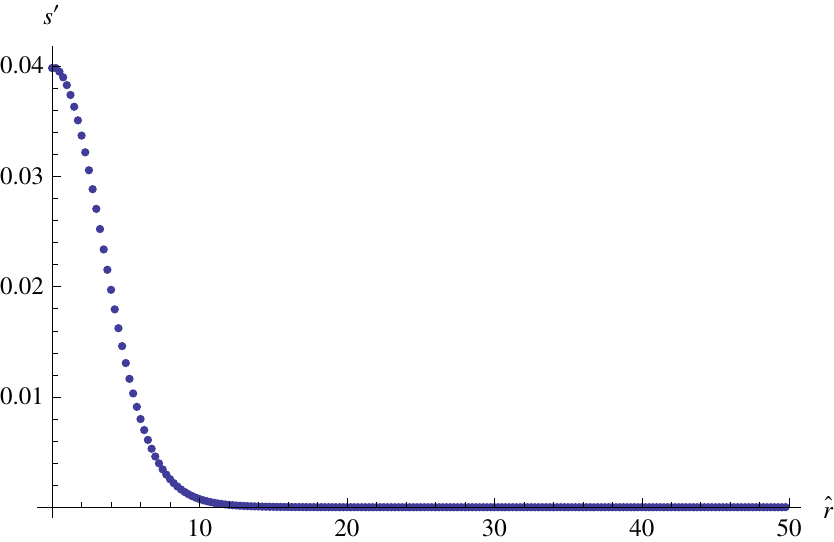}
\end{minipage}
\begin{minipage}{0.49\textwidth}\centering
 \includegraphics[scale=0.9]{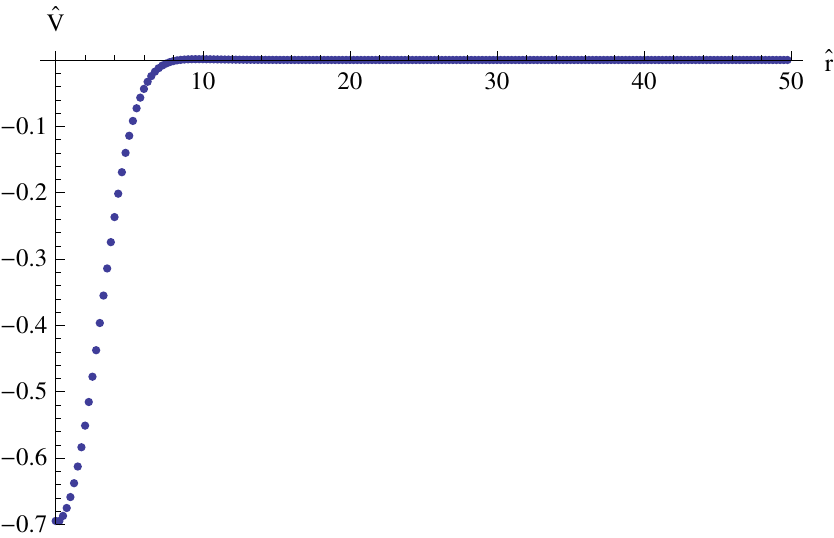}
\end{minipage}
\caption{\label{fig:7}Bounce configuration and potential corresponding to the decay towards the metastable vacua generated by diagonal spectators, for  $N_c=5,\t N_c=1,h=0.6,\kappa_\phi=7.7\cdot10^{-4}, \kappa_\lambda=38.5$.}
\end{figure}

As argued in \S\ref{subsec:numerics:dimensionless}, in the presence of spectator fields the decays of the ISS vacuum are expected to proceed dominantly towards the vacua involving nonzero VEVs for diagonal spectators. This section presents results concerning decays towards a vacuum in which all the spectator fields coupling to the diagonal of the meson field $X$ get VEVs, for $\t N_c=1$ and $N_c=5$. The fields that get different VEVs in the two vacua are $\chi,\tilde\chi,$ the diagonal components of $X$ and the spectator $S$, as well as $Y$ and  $\rho_5,\t\rho_5$. Again, boundary conditions motivate us to consider $\chi=\t\chi$ as well as $\rho=\t\rho$ along the bounce trajectory. Once more we take $X$ as diagonal, $X=x\mathbb{I}_5$, and the spectator field matrix is taken as $S={\rm diag}(s,s,s,s,-4s)$. To get canonical kinetic terms, the fields have to be rescaled as
\begin{align*}
 x=\frac{1}{\sqrt{10}}\,x', \quad Y=\frac{1}{2}\,Y',\quad \chi=\t\chi=\frac{1}{2}\,\chi',\quad \rho_5=\t\rho_5=\frac{1}{2}\rho',\quad s=\frac{1}{\sqrt{40}}\,s'.
\end{align*}
Thus, the bounce configurations involve 5 fields, $x',Y',\chi',r'$ and $s'$. The bounce action was computed for varying $h$ as well as for varying $\kappa_\phi$, keeping  the rest of parameters fixed; the results are shown in Figs.~\ref{fig:5} and \ref{fig:6}, respectively. Fig.~\ref{fig:7} shows one of the bounce configurations. From Figs.~\ref{fig:5}~and~\ref{fig:6} it is clear that one can obtain long-lived enough vacua with $S_b>400$ for order one values of $h$, which confirms the result quoted in ref.~\cite{SchaferNameki:2010iz}, which were obtained from a simpler calculation which ignored the tunneling along the $X$ and $Y$ directions. The more refined calculation presented here shows that the effect of $\kappa_\phi$ turns out to be very relevant, as shown in Fig.~\ref{fig:6}. The bounce action scales with $\kappa_\phi$ in a predominantly quadratic way. The results suggest then the dependence given in eq.~\eqref{eq:spectscaling}, valid for the choice of parameters of this section.

\section*{Acknowledgements}

This work has been financially supported by MICINN and the Fulbright Program through grant 2008-0800, and by the National Science Foundation under Grant No. PHY05-51164. The author wishes to thank Sebasti\'an Franco, Sakura Sch\"afer-Nameki and Gonzalo Torroba for valuable conversations, comments and suggestions.
\appendix
\section{One-loop Coleman-Weinberg potential}\label{app:VCW}

This paper presents computations of tunneling configurations using as potential the tree-level one plus the one-loop contribution to the effective potential, which is obtained as the Coleman-Weinberg potential $V_{\rm CW}$ evaluated for arbitrary background values of the fields. Using the supersymmetry preserving $\overline{\rm DR}$ scheme, the only contributions to $V_{\rm CW}$ come from integrating out the fields whose spectrum is not supersymmetric at tree-level. Since we are interested in the effective potential along the bounce trajectories interpolating between different vacua, we should take into account the fields that get a nonsupersymmetric spectrum not only in one of the vacua but also along the said trajectory.

In the ISS vacuum, the fields with a nonsupersymmetric spectrum are $\rho,\t\rho,Z$ and $\t Z$, while in the vacua associated with the vector spectators, these are a combination of $\chi,\t\chi$, some of the components of $\rho,\tilde\rho$ and off-diagonal components of $X$. Gauge fields should also be taken into account:  a crucial ingredient in the stabilization of all field directions in the ISS vacua is the gauging of part of the global symmetry, which generates D-term contributions that can break the degeneracy in the vector supermultiplet. This does not happen in the metastable vacua described in \S\ref{sec:vacua} but it may in the trajectories interpolating between them, so that new contributions to the one-loop potential might arise, in particular those yielding the mass terms that stabilize the pseudo-Goldstone modes of the ISS vacuum. These were modeled in ref.~\cite{SchaferNameki:2010iz} as a term of the form
\begin{align*}
 V\supset \frac{g^2}{16\pi^2}|\rho+\t\rho^*|^2|\chi+\t\chi^*|^2,
\end{align*}
where $g$ is the coupling of the gauged symmetry --note how these contributions vanish in both the ISS and spectator vacua, with $\rho=\t\rho=0$ and $\t\chi=\chi=0$, respectively. In the computations presented here these terms are not put by hand but will be encoded in the gauge field contributions to the Coleman-Weinberg potential; for simplicity a U(1) gauge symmetry is considered, under which only the fields $\rho,\t\rho$ are charged with charge $+1$ and $-1$, respectively.

In order to make the computation feasible by having analytic expressions for the mass matrices, it was decided to ignore the contributions from massive components of $X$ to the Coleman-Weinberg potential. This is only relevant when considering decays from the ISS vacuum to spectator vacua, but it is a good approximation due to the fact that the bounce configurations exit the barrier close to the ISS vacuum, in which the field components of the superfield $X$ have a supersymmetric tree-level mass matrix.

The one-loop Coleman-Weinberg potential at a scale $Q$ in the $\overline{\rm DR}$ scheme can be written as
\begin{align}\label{eq:VCW}
 V_{\rm CW}(Q^2)=\frac{1}{64\pi^2}\sum_n(-1)^{2s_n}(2s_n+1)m_n^4\Big(\log\frac{m_n^2}{Q^2}-\frac{3}{2}\Big)
\end{align}
where $s_n=1,1/2,0$ for vector, fermions and scalars, respectively, and $m_n$ are the corresponding field-dependent mass eigenvalues in a proper normalization of the fields. As justified in the discussion above, we include the following degrees of freedom:
\begin{itemize}
 \item Vector: U(1) gauge field,
\item Fermions: U(1) gaugino, fermionic components of the chiral fields $\rho,\t\rho,\t\chi,\chi,\t Z,Z$,
\item Scalars: Scalar components of the chiral fields $\rho,\t\rho,\t\chi,\chi,\t Z,Z$.
\end{itemize}
The mass eigenvalues in eq.~\eqref{eq:VCW} should be obtained for arbitrary background values of the fields with nontrivial tunneling profiles; by looking at the tree-level potential, it can be justified that these will be the fields taking nonzero VEVs in either of the vacua involved in the decay, since considering nonzero VEVs for additional fields will always increase the potential energy and drive the field configuration away from the sought-for critical point of the action. Thus, the mass matrices  of the fields mentioned above will depend on the background values of the fields $\chi,\t\chi,S,X,Y$ and $\rho,\t\rho$. Given the boundary conditions for the semiclassical considerations, one can restrict to configurations with  $\t\rho=\rho,\,\t\chi=\chi$. Furthermore, we will consider diagonal backgrounds for $X$ and $S$, and restrict to $N_c=5$, so that  $X=x\mathbb{I}_5$, $S={\rm diag}(s,s,s,s,-4s)$.
In order to write down the mass matrices, since we consider vacua in which one out of the $N_c$ components of $\rho,\tilde\rho$ gets a VEV, we separate the $\rho,\t\rho$ fields and the $Z,\t Z$ with which they mix in two groups, $\rho=(\rho_1,\rho_2)$, where $\rho_1$ is the component getting a VEV in the spectator vacua, and similarly for $\t \rho,Z,\t Z$. We then define the following multiplets:
\begin{align*}
&{\rm Vector}: b_\mu,\\
&{\rm Fermion}: \Psi=(\chi, \rho_1, Z_1, \t\chi,\t\rho_1, \t Z_1, \lambda), \quad \Psi'=(\rho_2, Z_2, \tilde \rho_2, \tilde Z_2), \\
&{\rm Scalars}:\,\, \t S_1={\rm Re}(\t\chi, \rho_1, Z_1,\chi, \t\rho_1,\t Z_1), \quad \t S_2={\rm Im}(\t\chi, \rho_1, Z_1,\chi, \t\rho_1,\t Z_1),\\
&\phantom{{\rm Scalars}:\,\,\, }\t S'_1={\rm Re}(\rho_2, Z_2, \t\rho_2,\t Z_2), \quad \t S_2={\rm Im}(\rho_2, Z_2, \t\rho_2,\t Z_2),
\end{align*}
where $b_\mu$ and $\lambda$ are the U(1) gauge field and gaugino. The mass matrices in the normalization consistent with eq.~\eqref{eq:VCW} are, for arbitrary background values of the fields $X,Y,\t\rho_1=\rho_1,\t\chi=\chi,s,x$, and for $\tilde N_c=1$,
\begin{align*}
 {\cal L}\supset \frac{1}{2}\,m^2_b b_\mu b^\mu+\frac{1}{2} \Psi^\intercal M_f \Psi+ \frac{1}{2} {\Psi'}^\intercal M'_f \Psi'+\frac{1}{2}\sum_{i=1,2}\big({\t S}_i^\intercal M^2_{\t S,i}\t S_i+\frac{1}{2}{{{\t S}_i}}^{'\intercal} M^2_{{\t S}',i}{\t S}'_i\big),
\end{align*}
\begin{align*}
 m^2_b&=4g^2\rho_1^2,\\
M_f&=\left[
\begin{array}{ccccccc}
 0 & 0 & h\rho_1 & hY & 0 & 0 & 0\\
0 & 0 & 0 & 0 & hx & h\chi & \sqrt{2}g\rho_1\\
h\rho_1 & 0 & 0 & 0 & h\chi & h^2\mu_\phi & 0 \\
h Y & 0 & 0 & 0 & 0 & h\rho_1 & 0\\
0 & h x & h\chi & 0 & 0 & 0 & -\sqrt{2}g\rho_1\\
0 & h\chi & h^2\mu_\phi& h\rho_1& 0 & 0 & 0\\
0 & \sqrt{2}g\rho_1 & 0 & 0 & -\sqrt{2}g\rho_1 & 0 & 0
\end{array}
\right],
\end{align*}
\begin{align*}
 M'_f=\left[
\begin{array}{c c c c}
 0 & 0 & hx & h\chi\\
 0 & 0 & h\chi & h^2\mu_\phi\\
 hx & h\chi & 0 & 0\\
h\chi & h^2\mu_\phi & 0 & 0 
\end{array}
\right],
\end{align*}
\begin{align*}
 (M^2_{S,1})_{ij}&=(M^2_{S,1})_{ji}\equiv m_{ij},\\
m_{11}&=m_{44}=h^2\rho_1^2+hY^2, & m_{12}&=m_{45}=2h^2\rho_1\chi,\\
m_{13}&=m_{46}=h^3\mu_\phi\rho_1 +h^2\rho_1 Y, & m_{14}&=h^2(\chi^2-\mu^2)+h^3\mu_\phi Y,\\
m_{15}&=m_{24}=0, & m_{16}&=m_{34}=h^2 x\rho_1,\\
m_{22}&=m_{55}=h^2\chi^2+h^2x^2+2g^2\rho_1^2, & m_{23}&=m_{56}=h^3\mu_\phi\chi+h^2\chi x,\\
m_{25}&=h^2(\rho_1^2-\mu^2)+h^3\mu_\phi x-2g^2\rho_1^2-4\lambda \Lambda h s, & m_{26}&=m_{35}=h^2 Y\chi,\\
m_{33}&=m_{66}=h^4\mu_\phi^2+h^2\rho_1^2+h^2\chi^2, & m_{36}&=m_{66}=0,
\end{align*}
\begin{align*}
 (M^2_{S,2})_{ij}&=(M^2_{S,2})_{ji}\equiv n_{ij},\\
n_{11}&=n_{44}=h^2\rho_1^2+hY^2, & n_{12}&=n_{45}=0,\\
n_{13}&=n_{46}=h^3\mu_\phi\rho_1 +h^2\rho_1 Y, & n_{14}&=h^2(\chi^2-\mu^2)+h^3\mu_\phi Y,\\
n_{15}&=n_{24}=0, & n_{16}&=n_{34}=h^2 x\rho_1,\\
n_{22}&=n_{55}=h^2\chi^2+h^2x^2, & n_{23}&=n_{56}=h^3\mu_\phi\chi+h^2\chi x,\\
n_{25}&=h^2(\rho_1^2-\mu^2)+h^3\mu_\phi x-4\lambda \Lambda h s,& n_{26}&=n_{35}=h^2 Y\chi,\\
n_{33}&=n_{66}=h^4\mu_\phi^2+h^2\rho_1^2+h^2\chi^2, & n_{36}&=n_{66}=0,
\end{align*}
\begin{align*}
 &{M'}^2_{S,1}={M'}^2_{S,2}=\\
&\left[
\begin{array}{cccc}
 h^2\chi^2+h^2 x^2\quad  & h^3\muphi \chi+h^2\chi x\quad & -h^2\mu^2+h^3\muphi x+h\lambda\Lambda s\quad& h^2 Y\chi\\
 h^3\muphi \chi+h^2\chi x\quad & h^4\muphi^2+h^2\chi^2\quad & h^2 Y\chi\quad &0\\
 -h^2\mu^2+h^3\muphi x+h\lambda\Lambda s\quad & h^2 Y\chi\quad & h^2\chi^2+h^2 x^2\quad & h^3\muphi \chi+h^2\chi x\\
h^2 Y\chi\quad & 0\quad  & h^3\muphi \chi+h^2\chi x\quad & h^4\muphi^2+h^2\chi^2
\end{array}
\right].
\end{align*}



\begin{thebibliography}{100}

\bibitem{Intriligator:2006dd}
  K.~A.~Intriligator, N.~Seiberg and D.~Shih,
  JHEP {\bf 0604}, 021 (2006)
  [arXiv:hep-th/0602239].

\bibitem{Essig:2008kz}
  R.~Essig, J.~-F.~Fortin, K.~Sinha, G.~Torroba, M.~J.~Strassler,
  JHEP {\bf 0903}, 043 (2009).
  [arXiv:0812.3213 [hep-th]].

\bibitem{Franco:2009wf}
  S.~Franco, S.~Kachru,
  Phys.\ Rev.\  {\bf D81}, 095020 (2010).
  [arXiv:0907.2689 [hep-th]].

\bibitem{Craig:2009hf}
  N.~Craig, R.~Essig, S.~Franco, S.~Kachru, G.~Torroba,
  Phys.\ Rev.\  {\bf D81}, 075015 (2010).
  [arXiv:0911.2467 [hep-ph]].

\bibitem{Behbahani:2010wh}
  S.~R.~Behbahani, N.~Craig, G.~Torroba,
  Phys.\ Rev.\  {\bf D83}, 015004 (2011).
  [arXiv:1009.2088 [hep-ph]].

\bibitem{SchaferNameki:2010iz}
  S.~Schafer-Nameki, C.~Tamarit, G.~Torroba,
  JHEP {\bf 1103}, 113 (2011).
  [arXiv:1005.0841 [hep-ph]].


\bibitem{Hook:2011ea}
  A.~Hook, G.~Torroba,
  [arXiv:1104.2331 [hep-ph]].

\bibitem{Green:2010ww}
  D.~Green, A.~Katz, Z.~Komargodski,
  Phys.\ Rev.\ Lett.\  {\bf 106}, 061801 (2011).
  [arXiv:1008.2215 [hep-th]].

\bibitem{Coleman:1977py}
  S.~R.~Coleman,
  Phys.\ Rev.\  {\bf D15}, 2929-2936 (1977).

\bibitem{Konstandin:2006nd}
  T.~Konstandin, S.~J.~Huber,
  JCAP {\bf 0606}, 021 (2006).
  [hep-ph/0603081].

\bibitem{Barnard:2010wk}
  J.~Barnard,
  JHEP {\bf 1101}, 101 (2011).
  [arXiv:1011.4944 [hep-ph]].

\bibitem{Coleman:1985ki}
  S.~R.~Coleman,
  Nucl.\ Phys.\  {\bf B262}, 263 (1985).

\bibitem{Benci:2010cs}
  V.~Benci, D.~Fortunato,
   [arXiv:1011.5044 [math-ph]].


\bibitem{Seiberg:1994pq}
  N.~Seiberg,
  ``Electric - magnetic duality in supersymmetric nonAbelian gauge theories,''
  Nucl.\ Phys.\  {\bf B435}, 129-146 (1995).
  [hep-th/9411149].

\bibitem{Nelson:1993nf}
  A.~E.~Nelson, N.~Seiberg,
  Nucl.\ Phys.\  {\bf B416}, 46-62 (1994).
  [hep-ph/9309299].

\bibitem{Callan:1977pt}
  C.~G.~Callan, Jr., S.~R.~Coleman,
  Phys.\ Rev.\  {\bf D16}, 1762-1768 (1977).


\bibitem{Frampton:1976kf}
  P.~H.~Frampton,
  Phys.\ Rev.\ Lett.\  {\bf 37}, 1378 (1976).

\bibitem{Weinberg:1994tk}
  E.~J.~Weinberg,
  [hep-ph/9406223].

\end{thebibliography}
\end{document}